\documentclass[pre,superscriptaddress,reprint]{revtex4-1}

\usepackage{graphicx}
\usepackage{color}
\usepackage{tabularx}
\usepackage{hyperref}

\begin{document}

\title{Group roles in unstructured labs show inequitable gender divide}

\author{Katherine N. Quinn}
\affiliation{Center for the Physics of Biological Function, Princeton University, Princeton NJ 08540, United States}
\affiliation{Initiative for the Theoretical Sciences, CUNY Graduate Center, New York NY 10016, United States}
\author{Michelle M. Kelley}
\affiliation{Laboratory of Atomic and Solid State Physics, Department of Physics, Cornell University, Ithaca, NY 14853, United States.}
\author{Kathryn L. McGill}
\affiliation{Department of Physics, University of Florida, Gainesville, FL 32611, United States.}
\author{Emily M. Smith}
\affiliation{Department of Physics, Colorado School of Mines, Golden, CO 80401, United States.}
\author{Zachary Whipps}
\affiliation{Department of Physics, Cornell University, Ithaca, NY 14853, United States.}
\author{N.G. Holmes}
\affiliation{Laboratory of Atomic and Solid State Physics, Department of Physics, Cornell University, Ithaca, NY 14853, United States.}

\date{\today}

\begin{abstract}
   Instructional labs are being transformed to better reflect authentic scientific practice, often by removing aspects of pedagogical structure to support student agency and decision-making. We explored how these changes impact men's and women's participation in group work associated with labs through clustering methods on the quantified behavior of students. We compared the group roles students take on in two different types of instructional settings; (1)~highly structured traditional labs, and (2)~less structured inquiry-based labs. Students working in groups in the inquiry-based (less structured) labs assumed different roles within their groups, however men and women systematically took on different roles and men behaved differently when in single- versus mixed-gender groups. We found no such systematic differences in role division among male and female students in the traditional (highly-structured) labs. Students in the inquiry-based labs were not overtly assigned these roles, indicating that the inequitable division of roles was not a result of explicit assignment. Our results highlight the importance of structuring equitable group dynamics in educational settings, as a gendered division of roles can emerge without active intervention. As the culture in physics evolves to remove systematic gender biases in the field, instructors in educational settings must not only remove explicitly biased aspects of curricula but also take active steps to ensure that potentially discriminatory aspects are not inadvertently reinforced.
\end{abstract}

\maketitle

The demographic composition of physicists is not representative of the general population, with men over-represented not only in number but also in high-ranking positions within the physics community~\cite{Pettersson2011}. In exploring the underlying mechanisms for this, there has been a large focus in education research on gaps in performance between men and women on concept inventories and course grades~\cite{Scherr,Madsen2013}. While informative, this approach provides an incomplete picture~\cite{Madsen2013,Andersson2016}; importantly, student persistence in physics can often be independent of their physics test scores~\cite{Sax2001}. New strides in science education research now include investigating more metrics such as sociocultural factors~\cite{Eddy2016,Rosa2016}, self-efficacy~\cite{Nissen2016,Kalender2019}, sense of belonging~\cite{Lewis2016}, and identity formation~\cite{Irving2016,Kalender2019,KalenderDec2019}. Moreover, participation in the physics community through the \textit{roles} people take on within the community can heavily shape one's identity as a physicist~\cite{Irving2015}. Within any community, members assume different roles as they take on different responsibilities, perform certain functions, and are perceived in specific ways by themselves and by the group~\cite{Wenger2002}. Understanding what roles develop throughout students' physics education is critical, as the field of physics is associated with masculinity, suggesting that a gendered division of roles may greatly influence the modern practice of physics.~\cite{Gonsalves2016,Francis2017}. 

Students have little direct experience with the field, however~\cite{Li2012}, and their perceptions of the field and their physics identities are developed through their immersion in physics courses. Many courses (including labs) involve significant group work, which can leverage the fact that strong peer relationships can benefit students' development of their science identities~\cite{stake2005,Close2016,Lock2015,Hazari2017}.
As with group work in other aspects of physics courses (such as cooperative problem solving, tutorial, or in-class lecture activities), lab activities require coordination of group members as they collect and interpret a common data set. Lab activities are distinct from other learning environments in that there are multiple distinct activities that must be carried out, so division of labor, and thus assigning distinct roles, is much more common. 

In this study, we explored patterns in the behaviors students exhibit in the context of physics labs. In doing so, we aim to better understand the group roles that emerge in these spaces. Labs provide an environment where students interact with peers and engage in physics experiments in ways that can influence their perception of physics and of themselves as physicists~\cite{Danielsson2009}. Furthermore, labs are changing nationally in response to calls to provide students with more authentic science experiences~\cite{AAPT}. Understanding how students behave and interact with each other in different lab environments, and the roles students assume in these settings, can inform educators and researchers when designing new pedagogy to better address inequities.

Identity formation is a complicated, multi-dimensional process that includes gender, race, physical ability, socioeconomic status, sexual orientation, and religion, among many, many other factors. The formative process includes individual agency as well as broader cultural and societal factors~\cite{Brickhouse2001}: the impact of the broader culture outside of the physics classroom strongly influences one's identity formation (such as a culturally-perceived notion of physics as a masculine field~\cite{Archer2016}). Importantly, how one develops a sense of identity impacts the set of available roles one may take on in a particular context, and strongly determines persistence in a particular field~\cite{Carlone2007}. In this study, we analyzed the quantified behavior profiles (discussed further in Sec.~\ref{sec:Methods}) as a way of probing the roles students take on in physics labs to understand some of the ways in which these roles can be equitably or inequitably divided.

We define an equitable division of roles as one in which all members are equally likely to assume each role, \textit{i.e.}, every role is available to every member. Note that this is different from equal or identical roles, in which every student performs the same function and thus would behave similarly from each other. An inequitable division of roles is one in which not every role is available to every member. For example, certain members are expected to assume, or prevented from taking on, certain roles. At the individual level, roles that are divided among group members may not be indicative of inequitable division. However, if students systematically behave differently in groups, then broader statistical analyses will reveal these overarching inequities. For instance, roles may be \textit{gendered}, in the sense that there is an inequitable gender divide, with men and women taking on systematically different roles~\cite{Butler1999,Doucette_2020}. 

Prior research has found that group work often involves inequitable participation between men and women. For example, female students participated less in group discussion when they were outnumbered by male students~\cite{PatHeller1992,adams2002observations} and responded disproportionately less than male students to instructor-posed questions in lecture~\cite{Eddy2014}. In physics lab courses, students have described the available roles themselves as being either masculine or feminine~\cite{Danielsson2009} and women have been found to engage less frequently with hands-on equipment~\cite{Javanovic1998,Holmes2014} or with computers~\cite{Day2016} when working in mixed-gender pairs. In contrast, contradicting results were found when comparing the performance of male-majority, female-majority, and mixed groups on engineering design tasks across two different courses~\cite{Laeser2003}. 

Individual students' behaviors can be used to probe the roles they take on in physics labs, and are likely a result of their personal identity (gender or otherwise), the particular instructional context, and the broader physics culture~\cite{Butler1999, Traxler2016, Gosling2017, Carlone2007, Day2016, Jovanovic1998}. Understanding students' experiences in labs through the behaviors and roles they take on both highlights existing gender disparities as well as informs future research on students' persistence in science. We specifically sought to understand the impact of \textit{different} instructional lab environments on student roles and how these roles are divided between men and women. One way to make labs more authentic is to make them discovery-based and inquiry-driven, removing structure from the lab. How does removing pedagogical structure in the lab impact these learning environments? Specifically, what impact does pedagogical structure have on the equitable division of roles within groups?

\section{Materials and Methods}
\label{sec:Methods}

All participants in this study were undergraduate students at a major research university enrolled in the honors-level mechanics course of a calculus-based physics sequence. The course was designed for physics majors and open to students across the sciences and engineering. The sample of prospective physics majors is an important population for this study, given the potential link between students' experiences, roles, identity, and persistence in physics~\cite{Carlone2007}. We explored students' behaviors in two different types of lab instruction. 

The highly-structured \textit{traditional labs} were designed to reinforce physics content knowledge presented in lecture. Students were provided with detailed paper worksheets to follow during lab, guiding them through experiments that provided them with hands-on experience. The lab guides provided explicit details about what and how much data to collect and posed targeted conceptual physics questions to support making predictions and interpreting results. Students worked in groups to collect data for the experiments and submitted individual paper worksheets.

In contrast, the less structured \textit{inquiry labs} were designed to emphasize the process of experimentation in physics (see, for example, Ref.~\cite{Holmes2015, SmithPRX, HolmesTPT2019, Holmes2020}). Students were provided with a specific goal, but were expected to design their own experiment to achieve that goal. Lab guides prompted students to design data collection methods to reflect on results, and to design follow-up investigations to improve or extend their investigations. Students worked collaboratively to design and implement their experiments and submitted only one electronic notebook as a group. Reference~\cite{SmithPRX} includes additional detail about the differences between the conditions, including differences between students' learning engagement with experimentation, and attitudes towards experimental physics.

The same mechanics course was taught twice during the academic year, once in the fall semester and then again in the spring semester. Students from both semesters were included in this study. During the first semester, all students attended the same lecture, mixed together in discussion sections, but were separated into two pedagogically different lab types discussed below (three \textit{traditional lab} sections and two \textit{inquiry lab} sections). During the second semester, the two lab sections under study were both inquiry labs. Note that we observed students across multiple lab periods throughout the course of the semester (and each student appeared in only one semester), and so while each student is in one lab section they appear in multiple lab periods. All participants were unaware of the differences between lab types: students in the first semester self-selected into their lab sections prior to the start of the course by registering for the course, and only the inquiry lab sections were available to students in the second semester. Student groups varied every period, and were randomly assigned.

The role a student takes on in their group is a highly complex reflection of the function they serve in the group, and depends on numerous factors from the individual to the cultural level. Because this study explores student roles in physics labs, we assume that these roles are in some way correlated with their behavior in these labs, such as handling of equipment or of computer usage. To probe the roles that students assumed in physics labs, we analyzed the quantified behavior profiles of 143 students across multiple lab periods. We collected data for this study at two levels of granularity.

First, \textit{coarse behaviors} were captured at five minute intervals for all students in multiple lab periods. The codes were determined by what the students were handling: (1)~lab desktop computer, (2)~personal laptop or other device, (3)~writing on paper, (4)~handling equipment, or (5)~engaging in some other activity. We used the Other code to capture all other behaviors such as: discussing within their group, with another group, or with the instructor; engaging in whole-class discussions; writing on whiteboards; or engaging in off-task behaviors. Note that the Other code was constructed to ensure all time was coded for every student, and therefore captures many different behaviors. The choice of codes were designed to capture enough detailed information as possible about every student, coded in real time, while reflecting the lack of \textit{a priori} knowledge of what the exact group roles were. The behaviors of each student in each lab period were amassed to create a \textit{profile} of their behaviors during that lab period. Unfortunately, given the observation protocol (discussed in detail in Sec.~\ref{sec:coarsebehaviors}) where each student was observed over the course of an entire lab period, subdividing the Other code could not be done quickly enough and with enough accuracy by the researchers. Instead, a second analysis of such detailed behavior was performed using video from single groups, and discussed in greater detail in Sec.~\ref{sec:finebehavior}. 

\subsection{Collecting Demographic Information}

We used in-class surveys to obtain student demographic information. In all, 143 students across multiple lab sections were used in this study. While they had the option to disclose a gender other than \textit{woman} or \textit{man}, no student chose to do so, and only two students did not disclose their gender identity. As a result, all students were included in the initial cluster analysis, however the gender analysis follows the traditional gender binary of \textit{woman} or \textit{man} (with the two undisclosed students omitted from the graphs in Fig.~\ref{fig:clusters} and Fig.~\ref{fig:clusterDistribution} due to insufficient statistics). Table~\ref{table:Demographics} shows the demographic breakdown of student participants in this study. To obtain the standard error on the fraction of a population (such as in Table~\ref{table:Demographics} or Fig.~\ref{fig:clusterDistribution}), we used the following:
\begin{eqnarray}
\label{eq:PopErr}
\texttt{Err}(p,N) = \sqrt{\frac{p(1-p)}{N}}
\end{eqnarray}
where $p$ is the fraction of the population, and $N$ is the size of the total population.

\begin{table}[htbp]
\caption{\textbf{Student demographics} of this study. Errors were computed using standard error for population fractions, shown in Eq.~\ref{eq:PopErr}. In all, 143 students were considered in this study. \label{table:Demographics}}
\begin{ruledtabular}
\begin{tabular}{l c c c c}
 & \multicolumn{2}{c}{\textbf{Traditional Labs}} & \multicolumn{2}{c}{\textbf{Inquiry Labs}} \\
					& $N$ & \% & $N$ & \% \\
 \hline
Women & 11 & $19 \pm 5$ & 21 & $25\pm 5$ \\
Men & 46 & $79\pm 5$ & 63 &  $74 \pm 5$ \\
Undisclosed & 1 & $2\pm2$ & 1 & $1\pm 1$
\end{tabular}
\end{ruledtabular}
\end{table}

\subsection{Quantifying Coarse Student behaviors}
\label{sec:coarsebehaviors}

In all lab sections, observers documented student behaviors following the observation protocol used in Ref.~\cite{Day2016}. Every five minutes, an observer noted each student's actions in the lab using one of five codes: Desktop, Equipment, Laptop, Paper, and Other. One code was applied to each student in the class at each five-minute interval, except in cases where students could not be observed (e.g. because they were late or left early). The codes are described in Table~\ref{table:codes}, and were based on what a student could be handling in the lab. The Other code captured all other behaviors such as engaging in whole-class discussions, writing on whiteboards, discussing with the TA or UTA, and off-task behaviors, ensuring that all in-lab time was coded. The Desktop code was separated from the Laptop code because the Desktop was often required for data collection (e.g. because it was directly connected to a detector or piece of equipment). Furthermore, desktops were shared within groups whereas the Laptop code was ascribed to students handling personal devices. While desktops were present in both lab types, only students in the inquiry labs actively used laptops to analyze data, document their lab procedures and submit their electronic notebooks. 

\begin{table}[htbp]
\caption{\textbf{Action codes used in observations}. The Laptop code is used for both handling a laptop or personal device (students used laptops, phones, and tablets for the purpose of notetaking, writeup, data analysis and reading instructions in the inquiry labs). \label{table:codes}}
\begin{tabular}{ll}
\hline\hline
\textbf{Code} & \textbf{Description} \\
 \hline
Desktop & Using the desktop computer at the lab bench. \\
Equipment & Handling equipment. \\
Laptop & Using a laptop or personal device. \\
Paper & Writing on paper or in a notebook. \\
Other & Other action or behavior.\\
\hline\hline
\end{tabular}
\end{table}

The codes used in this study, in particular the Other code, are very coarse and so multiple behaviors can fall under the same code (\textit{e.g.} the Laptop code includes using a laptop for data anlysis as well as for note taking, the Other code captures activities such as discussing the lab with group members or engaging in off-task talking with group members). Given the observation protocol, it was not possible for an observer to differentiate between these different and more nuanced behaviors in real time for every student, and so a second analysis was performed, and the details of which are outlined in the Sec.~\ref{sec:finebehavior}.

To validate our observation procedure, two observers coded student actions in the same lab period using the described protocol but at different five-minute intervals. If we had had each observer code the same student at the same time, we would have only evaluated the reliability of the codes. Instead, observers were specifically not coding the same student at the same time. Thus, comparing the overall code count for each student provides a measure of reliability of the codes \textit{recorded at five-minute intervals}. By comparing the overall code count for each student, we provide a measure of reliability about the overall method. This method limits us, however, from comparing individual student behavior over time in the lab period. Thus, all analysis is performed on the student profiles, which aggregate their behaviors throughout the lab period. Note that because observers were explicitly not observing the same student at the same time, percent agreement or calculating Cohen's Kappa would not provide the necessary information to validate the method. Instead, a standard chi-squared analysis was performed on the contingency table constructed from the accumulated codes (the frequency each observer noted each code, summed over all students). We used the criteria that if two sets of observations are statistically indistinguishable from each other, then the observers captured the same overall profiles for the students in the lab session. Note that, if either (1)~there was not agreement between the codes, or (2)~the five-minute interval did not accurately capture student behavior when averaged over a lab period, then there would be disagreement in these overall distributions.

\begin{figure}
    \centering
    \includegraphics[width=0.45\textwidth]{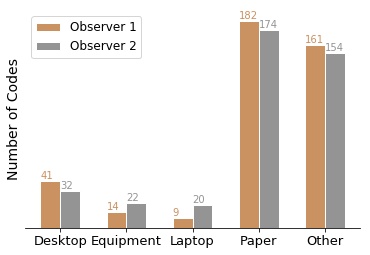}
    \caption{\textbf{Bar plot of code counts from two observers} used to form the basis of a chi-squared test to validate the observation protocol used in this study. Two observers documented the same lab period, and the resulting contingency table (given by the raw counts displayed on the graph and shown in Table~\ref{table:contingency}) was used to determine statistical validity of the method. Here, the two distributions are statistically indistinguishable indicating that the observers captured the same distribution of student actions.}
    \label{fig:interrater}
\end{figure}

In all cases observers' distributions were statistically indistinguishable, and so single observers coded subsequent lab periods. When attempts were made at subdividing the codes, for instance to capture students performing data analysis vs. notetaking or identifying if group discussions were off task, we were not able to obtain agreement between observers. As such, we used the protocol detailed in this section. We provide an example of observer comparisons for illustrative purposes. A sample graph of the accumulated codes for two observers in a traditional lab section is presented in Fig.~\ref{fig:interrater}. The contingency table constructed from these observations is given by Table~\ref{table:contingency}. Because the two distributions are statistically indistinguishable, the observers captured the same distribution of student actions. 

\begin{table}[htbp]
\caption{\textbf{Sample contingency table} used to determine if two distributions are statistically different. Two observers documented the same lab period, and a chi-squared test was performed to determine if the resulting distributions are statistically similar or dissimilar. Here, we obtain $p>0.1$, indicating that the observers captured the same distribution of student actions.\label{table:contingency}}
\begin{tabular}{l c c c c c}
\hline\hline
\textbf{Observer} & \textbf{Desktop} & \textbf{Equipment} & \textbf{Laptop} & \textbf{Paper} & \textbf{Other}\\
 \hline
1 & 41 & 14 & 9 & 182 & 161 \\
2 & 32 & 22 & 20 & 174 & 154\\
\hline\hline
\end{tabular}
\end{table}

Because students were observed during multiple lab periods over a full semester, we were able to document individual students more than once. As a result, we obtained 522 unique \textit{student profiles}, each quantifying the actions of one student in one lab period through the frequency of associated codes. Table~\ref{table:profileDemographics} shows a demographic breakdown of the student profiles used in this study.

\begin{table}[htbp]
\caption{\textbf{Demographic breakdown of student profiles} measured in this study. Errors were computed using standard error for population fractions, shown in Eq.~\ref{eq:PopErr}. In all, 143 students were observed across multiple lab periods, resulting in 522 unique student profiles. \label{table:profileDemographics}}
\begin{tabular}{l c c c c}
\hline\hline
 & \multicolumn{2}{c}{\textbf{Traditional Labs}} & \multicolumn{2}{c}{\textbf{Inquiry Labs}} \\
					& $N$ & \% & $N$ & \% \\
 \hline
Women & 34 & $18 \pm 3$ & 87 & $26\pm 2$ \\
Men & 152 & $81\pm 3$ & 246 &  $74 \pm 2$ \\
Undisclosed & 2 & $1\pm1$ & 1 & $0.3\pm 0.3$\\
\hline\hline
\end{tabular}
\end{table}

\subsection{Cluster Analysis}
\begin{figure}
    \centering
    \includegraphics[width=0.45\textwidth]{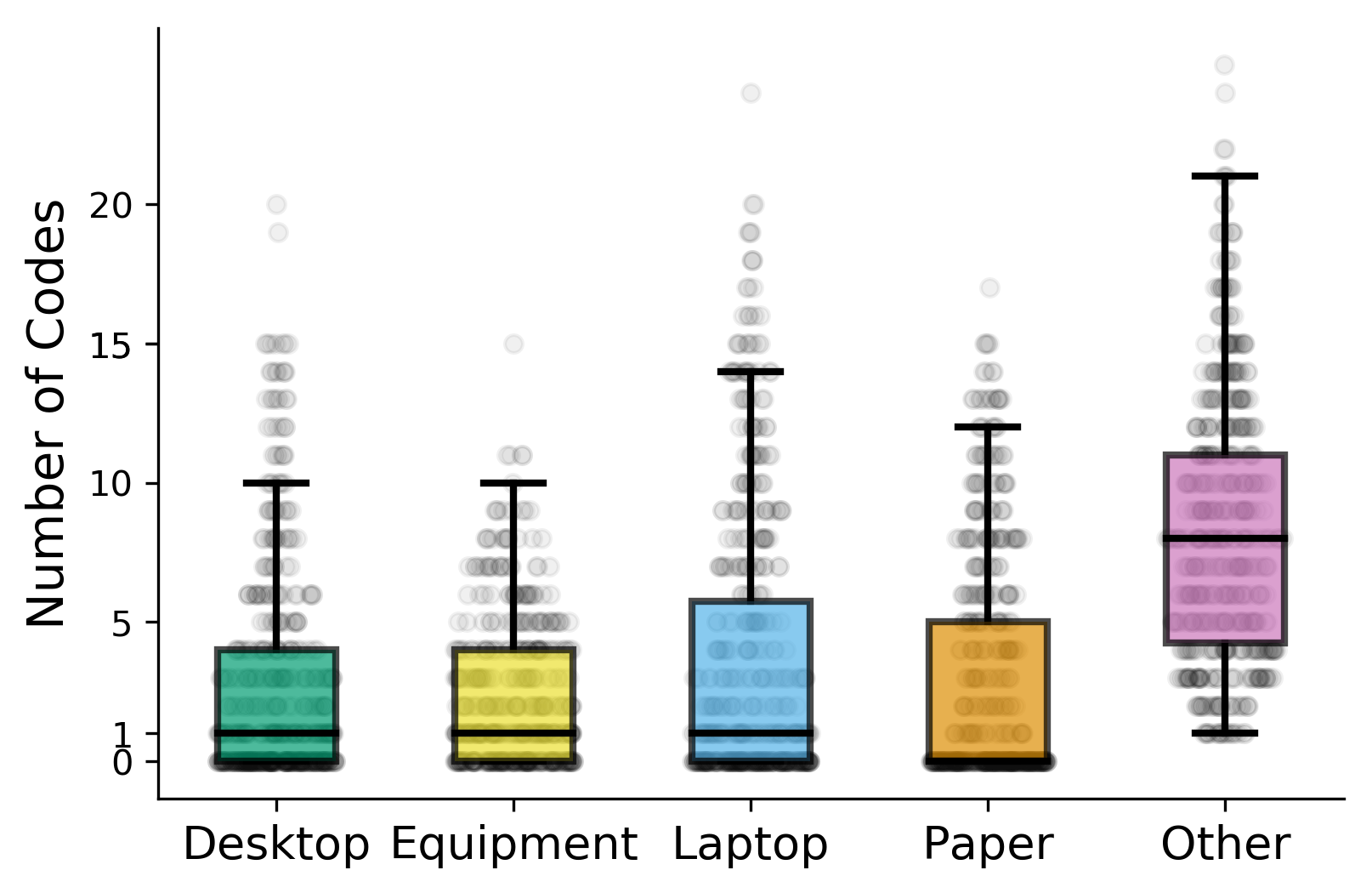}
    \caption{\textbf{Box plots of raw data} revealing the highly non-Gaussian nature of the code distributions. Each faded point is the accumulated codes for a student in a lab period for a particular category (the horizontal spread of the points is just to visualize all the points), and so darker regions represent more total codes of that value (with the darkest regions near zero). Note that the median for all codes except Other is less than or equal to one, reflecting the fact that over half of students were observed engaging in that behavior once or less than once. This, combined with the fact that there are a large number of outliers, is an indication that students either engage in a particular activity a lot or not at all.}
    \label{fig:box}
\end{figure}

The distribution of coarse behavior code frequencies are highly skewed, with most students engaging in a particular activity infrequently or not at all and some students engaging in an activity a lot. Figure~\ref{fig:box} shows box plots of the raw data, illustrating the non-Gaussian features of the data. For this reason, we performed a cluster analysis instead of methods that rely on the assumption of Gaussian distributions. Clustering can account for non-linearities missed in common regression analyses, capturing \textit{dominant} behavior as opposed to \textit{average} behavior, and has been used in similar studies of this type to provide fruitful results~\cite{Corpus2014}. By performing a demographic analysis on the student groupings (i.e. clusters) we can quantitatively characterize coarse gendered behavior.

To perform a cluster analysis on multidimensional data, the scales for each measure must be the same. In this study, there were two major effects present which caused differences in scales that we accounted for. First, the amount of coded time for each student was highly variable, ranging from less than 45 minutes to over 175 minutes. To account for this effect, we normalized each student profile. In this way, each measure represents the fraction of time spend on a particular task. Second, there is the inherent differences in the five measures. For instance, from Fig.~\ref{fig:box}, we can see that the distributions for Other is more spread out than for Equipment. To account for this, each measure was grand mean scaled so that, averaged over all students, each measure had mean 0 and standard deviation of 1. In doing so, each measure becomes a Z-score~\cite{Corpus2014,Schmidt2017}. Thus, each student's Z-score tells us whether the time they spent on a particular activity was above or below average as compared to other students. Moreover, the Euclidean distance between two profiles has a statistical interpretation in this Z-score format: it measures the dissimilarity of two student profiles in units of standard deviations~\cite{Corpus2014}.

We performed a standard k-means clustering on the rescaled student profiles. K-means is an iterative algorithm, where the researcher specifies the number of clusters. The algorithm clusters and then re-clusters the data in an iterative manner until the sum of the square of the distances from all points to their respective cluster's center is minimized and no point changes cluster between iterations~\cite{kMeans}. 

\begin{figure}
    \centering
    \includegraphics[width=0.45\textwidth]{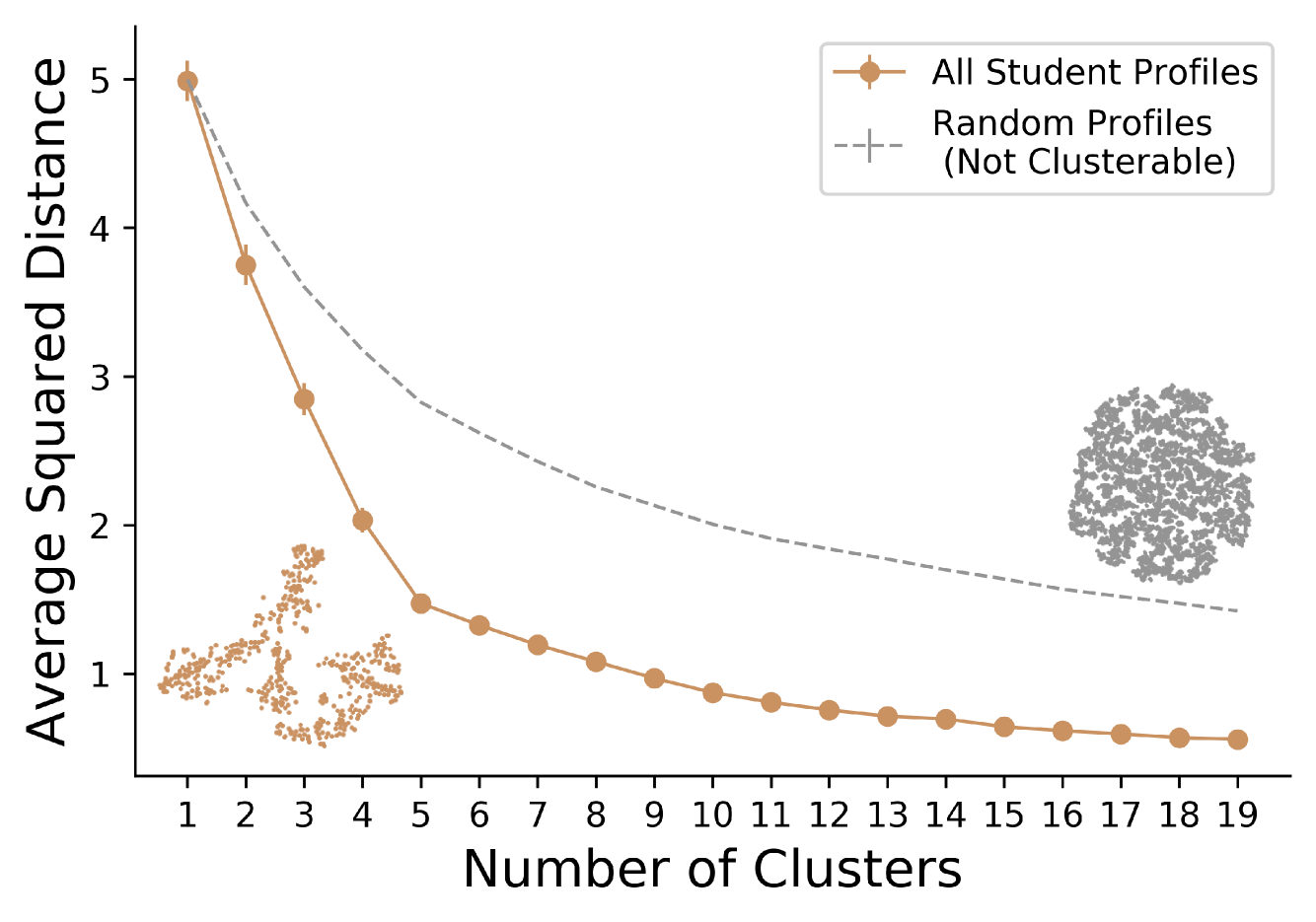}
    \caption{\textbf{Elbow plot} used to determine the optimal number of clusters for the data. The average squared distance from each point to the center of its assigned cluster is plotted as a function of the number of clusters. There is a kink at five, indicating that the optimal number of clusters for the data is five. Our results were compared against 10,000 randomly generated student profiles. Note that the elbow is well below random, a sign that the data can be clustered. Superimposed on the graph is a two-dimensional visualization of the data and random points for qualitative comparison. The data show structure (brown points in lower left), whereas the random points form a blob (grey points in center right).}
    \label{fig:elbow}
\end{figure}

Note that not all data can be meaningfully clustered. For example, even if all data form a structure-less blob, a researcher can still input two or more clusters and the algorithm will converge to a solution. Therefore, in order to determine (1)~if the data are clusterable, and (2)~if so, what the optimal number of clusters is, we used the elbow method~\cite{elbow}. We plotted the average squared distance from each point to the center of its assigned cluster, as a function of the number of clusters, and compared the results to 10,000 randomly generated student profiles. We used enough random data to numerically generate a smooth function and ensure that the comparison is not hindered by statistical fluctuations. The results of the elbow plot are shown in Fig.~\ref{fig:elbow}. The plot for our collected data was substantially below random, indicating that the data is clusterable. There is a distinct kink in the plot for five clusters, indicating that the optimal number of clusters is five. 

From the elbow plot in Fig.~\ref{fig:elbow}, specifically from looking at the drop in average squared distance from each point to the center of its cluster for five clusters compared to one, we can see that the five optimal clusters account for 70\% of the variance in the data. By looking at the distances confined to each of the five measures (i.e. generating similar figures as that of Fig.~\ref{fig:elbow} for each measure, where the max value would be one instead of five), we found that the five optimal clusters account for 73\% of Desktop use, 60\% of Equipment use, 78\% of Laptop use, and 59\% of Other activities). This is well above the 50\% threshold used for a study of this type~\cite{Corpus2014,Schmidt2017}. 

\begin{figure*}
    \centering
    \includegraphics[width=0.95\textwidth]{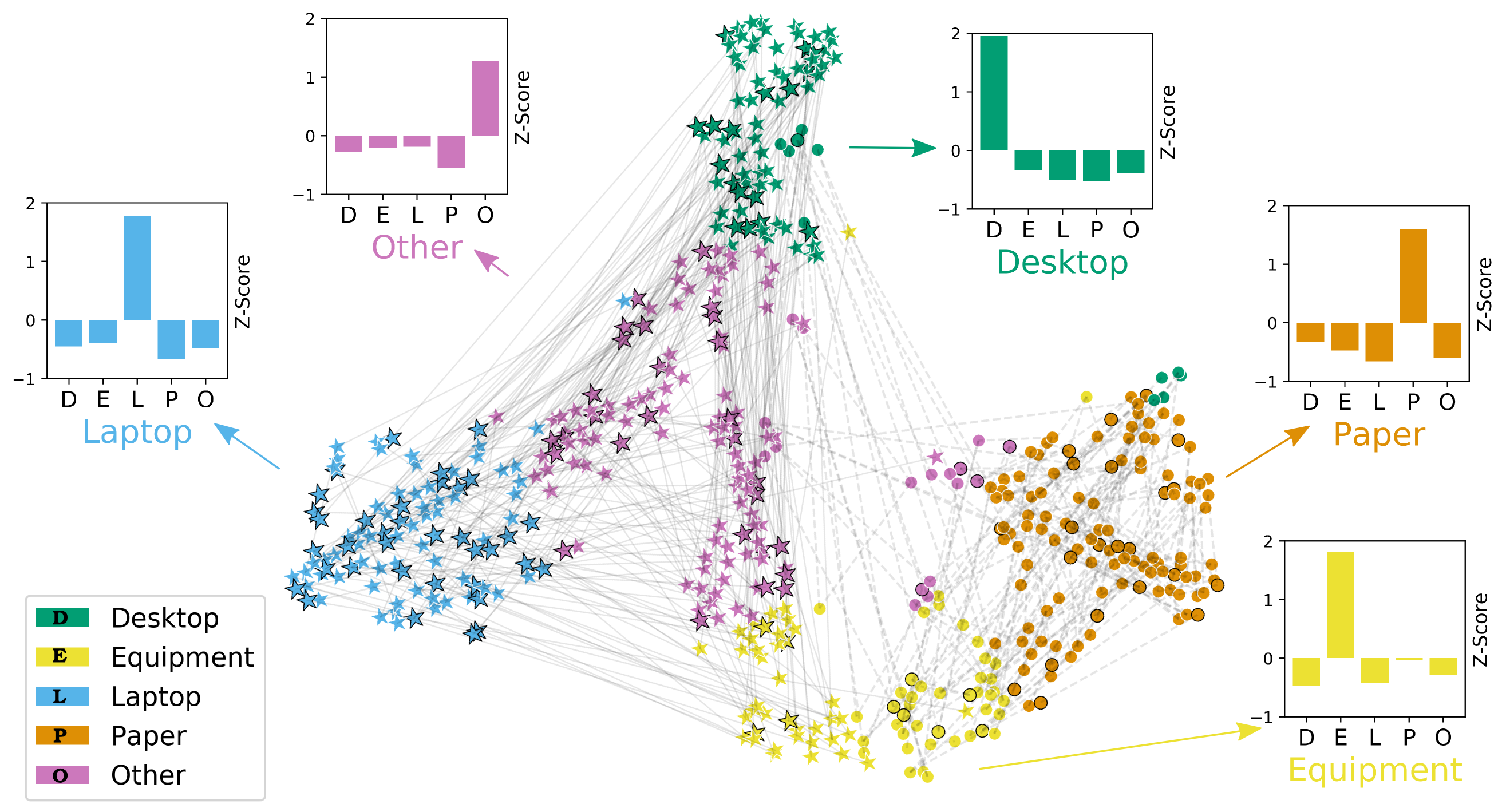}
    \caption{\textbf{Two-dimensional visualization of behavior clusters} and their centers. Each point represents a unique student profile, with profiles from the same group connected by a grey line (solid for less-structured inquiry labs, and dashed for highly-structured traditional labs). Circles represent students in the traditional labs and stars in the inquiry labs, and black edges indicate women's profiles. All points in the Laptop cluster are stars, whereas all points in the Paper cluster are circles, a reflection of the pedagogical differences in the labs (students in the traditional labs were filling out paper worksheets, whereas in the inquiry labs were filling out electronic notebooks). Clusters are characterized by their centers, and here the centers of the five clusters are given by large Z-scores for each of our codes.}
    \label{fig:clusters}
\end{figure*}

We provide a 2D visualization of the clusters using t-SNE~\cite{VanDerMaaten2008}, with each dot representing a profile colored by its assigned cluster (Fig.~\ref{fig:clusters}). Figure~\ref{fig:clusters} is a two-dimensional representation of a five-dimensional space, and so is used primarily for qualitative illustration.

Clusters from k-means are characterized by their centers. Here, the centers of the five clusters matched the five codes used in this study and so we labelled the clusters accordingly. Therefore, the clusters characterize ``high users'' of a particular measure. Note that this description fits with the raw data, shown in Fig.~\ref{fig:box}, which illustrates that the majority of students engage in a particular task either frequently or very rarely. For example, students in the yellow cluster of Fig.~\ref{fig:clusters} spent a larger fraction of their time on the equipment than the average student, so this cluster is referred to as the Equipment cluster. This is a feature of student behaviors, and not due to the number of codes used in this study. For instance, one could imagine a scenario in which all students behave nearly identically, with minor differences described by fluctuations: in that case, the data would form a five-dimensional Gaussian cloud centered at zero, and an elbow plot that matches random noise. Or, one could imagine a scenario in which only students handling equipment handle the lab desktop, in which case a cluster would emerge that couples the two respective codes.

We used the clusters that emerged from the data to coarsely characterize the roles students take on in labs. Generally, roles within groups are complex and multidimensional and could be further explored in greater detail through more detailed video analysis (discussed in Sec.~\ref{sec:finebehavior}), student interviews, or anthropological investigations. The analysis performed here provides a coarse-grained perspective on the division of roles within groups, and will ultimately reveal the unexpected inequities in role divisions (discussed next).

Because each student had multiple profiles, arising from several lab periods over the course of a semester, we investigated whether or not it is possible to further collapse the profiles to determine ``semester-long'' behaviors. We did this by analyzing whether or not individual students' profiles appear in multiple clusters over the course of a semester. In the traditional labs, \smash{$87\pm4\%$} of students have profiles appearing in more than one cluster. Similarly, \smash{$86\pm 4\%$} of students in the inquiry lab appear in more than one cluster. Because so many students have profiles appearing in multiple clusters, the weekly variation in an individual's profile is too great to further collapse (for numerous reasons, such as variability in lab content and students changing lab partners).

\subsection{Describing Detailed Student behavior}
\label{sec:finebehavior}

We used video recording of single-groups during full lab periods to better describe student behavior in more detail than captured in the previous section. In all, ten videos were coded, decomposing 23 profiles from 17 students (five students appeared in more than one video). BORIS software was used to code videos~\cite{BORIS}, specifically the fraction of time students engaged in different behaviors. 

\begin{figure*}
    \centering
    \includegraphics[width=0.8\textwidth]{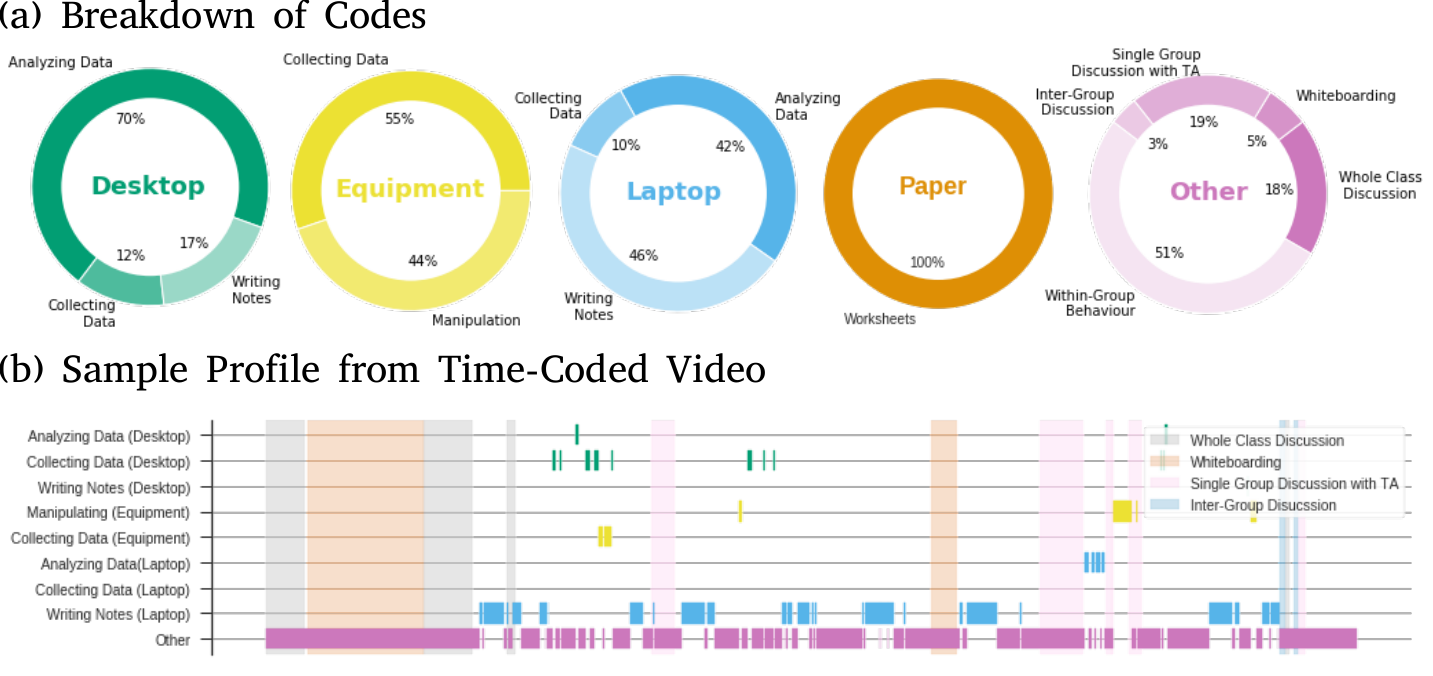}
    \caption{\textbf{Breakdown of codes} by decomposing coarse behavior (e.g. ``handling laptop'') into more fine-grained behavior (e.g. ``analyzing data''). Ten videos were coded, resulting in 23 decomposed profiles from 17 different students (five students appeared in more than one video). (a)~A breakdown of each code, showing the fraction of time students engaged in a particular task while coded as a particular behavior. Three of the five codes (Desktop, Equipment, and Laptop) were directly decomposed into sub-codes while analyzing videos, as shown in (b)~illustrating a sample coded time-series. Four additional ``group states'' were coded in the videos, representing large group behavior (discussing with a TA or UTA, conversing with other groups, whole class discussions and announcements, and using a whiteboard). We decomposed the Other code by overlapping it with these larger group states. The Paper code was purely represented by students filling out paper worksheets in the traditional labs.}
    \label{fig:CodeBreakdown}
\end{figure*}

The five codes in Table~\ref{table:codes} were further broken down by what a student was doing (e.g. analyzing data) while engaged in that coarse behavior (e.g. using the Desktop) as shown in Fig.~\ref{fig:CodeBreakdown}. The Paper code was used to predominantly describe students filling out paper worksheets in the traditional labs, and so it was not further decomposed. Students in the inquiry labs predominantly used whiteboards for calculations, and very rarely used paper. Both the Desktop and Laptop codes were used to describe students analyzing data, collecting data, or writing lab notes, and so both of these codes were broken down in this way. However, when collecting data, the desktop was often connected directly to equipment whereas gathering data on a laptop was purely represented by students manually entering data into their electronic notebook or analysis software. Students handling equipment were primarily doing so to either collect data or manipulate the setup in some way (setup, cleanup, calibration, playing) and so the Equipment code can be further decomposed into these two tasks. In this way, the Desktop, Equipment, Laptop, and Paper codes were explicitly decomposed.

To better describe student behavior while coded as Other, we introduced four new state codes. These were used to describe significant events in lab, and are elaborated in Table~\ref{table:videoCodes}. By overlapping the event codes with Other, we broke down the Other code and provide a more qualitative picture of classroom activities, such as engaging in whole-class discussions, using whiteboards to sketch out ideas and concepts, single group discussions with the TA or UTA, or engaging in inter-group discussions with neighboring groups.

\begin{table}[htbp]
\caption{\textbf{Event codes used in video observations}. These codes described significant events in the lab, and were used to decompose the more coarse-grained Other code. A sample time series illustrating a coded video is shown in Fig.~\ref{fig:CodeBreakdown}{(b)} \label{table:videoCodes}}
\begin{tabularx}{0.45\textwidth}{p{3cm}X}
\hline\hline
\textbf{Code} & \textbf{Description} \\
 \hline
Whole Class Discussion & The TA or UTA makes an announcement to the class, or holds a whole class discussion. \\ \hline
Whiteboarding & Students perform invention activities in the lab, and use a white board to sketch out ideas and concepts. \\ \hline
Single Group Discussion with the TA & TA or UTA engages in a discussion with the group (but not as part of a whole class discussion).  \\ \hline
Inter-Group Discussion & Groups compare results or discuss among each other (not as part of a whole class discussion).\\
\hline\hline
\end{tabularx}
\end{table}

To validate this method, two observers coded the same video as a means of testing the inter-rater reliability. The level of agreement was assessed with Cohen's Kappa where a value of \smash{0.61--0.80} represents substantial agreement. Two observers coded the same video, and obtained a Cohen's Kappa value of 0.79, indicating substantial agreement between the two. As a result, only one researcher coded the subsequent videos.

Video analysis was also used to better understand task allocation. Point-events were identified when one student explicitly instructed another to perform a task. We broke down the criteria for inclusion as a point event and exclusion as a point event in the following way:
\begin{itemize}
    \item \textit{Criteria for Inclusion:} A student needs to be addressing another, and explicitly direct them in some way, such as by saying ``you should do X''.
    \item \textit{Criteria for Exclusion:} Suggesting a task should be done that a student assumes without being asked is not included. Examples of such events are characterized by statements such as ``We should do X.'', ``I think we should focus on X.'', ``Does someone want to work on X?''. Additionally, a student asking another for help performing a task is excluded (such as asking another student how to sum a row in a spreadsheet, and the student telling them how).
\end{itemize}
In total, we found eight point events for inclusion from all ten videos. All such events were quick, directed comments related to a task the student was already engaging in. Therefore, as described in the main text, we conclude that no tasks were explicitly assigned by another student.

\section{Results}
\subsection{Identifying course-wide behavior patterns through cluster analysis}

We analyzed the demographic composition of each behavior cluster by lab type (highly-structured traditional or less-structured inquiry-based), gender (students' self-reported gender identity of \textit{man} or \textit{woman}), and group composition (mixed-gender or single-gender groups). In all cases, when comparing the composition of behavior clusters, we used a chi-squared test of frequencies on the contingency tables of the raw counts.

\begin{figure*}
    \centering
    \includegraphics[width=0.75\textwidth]{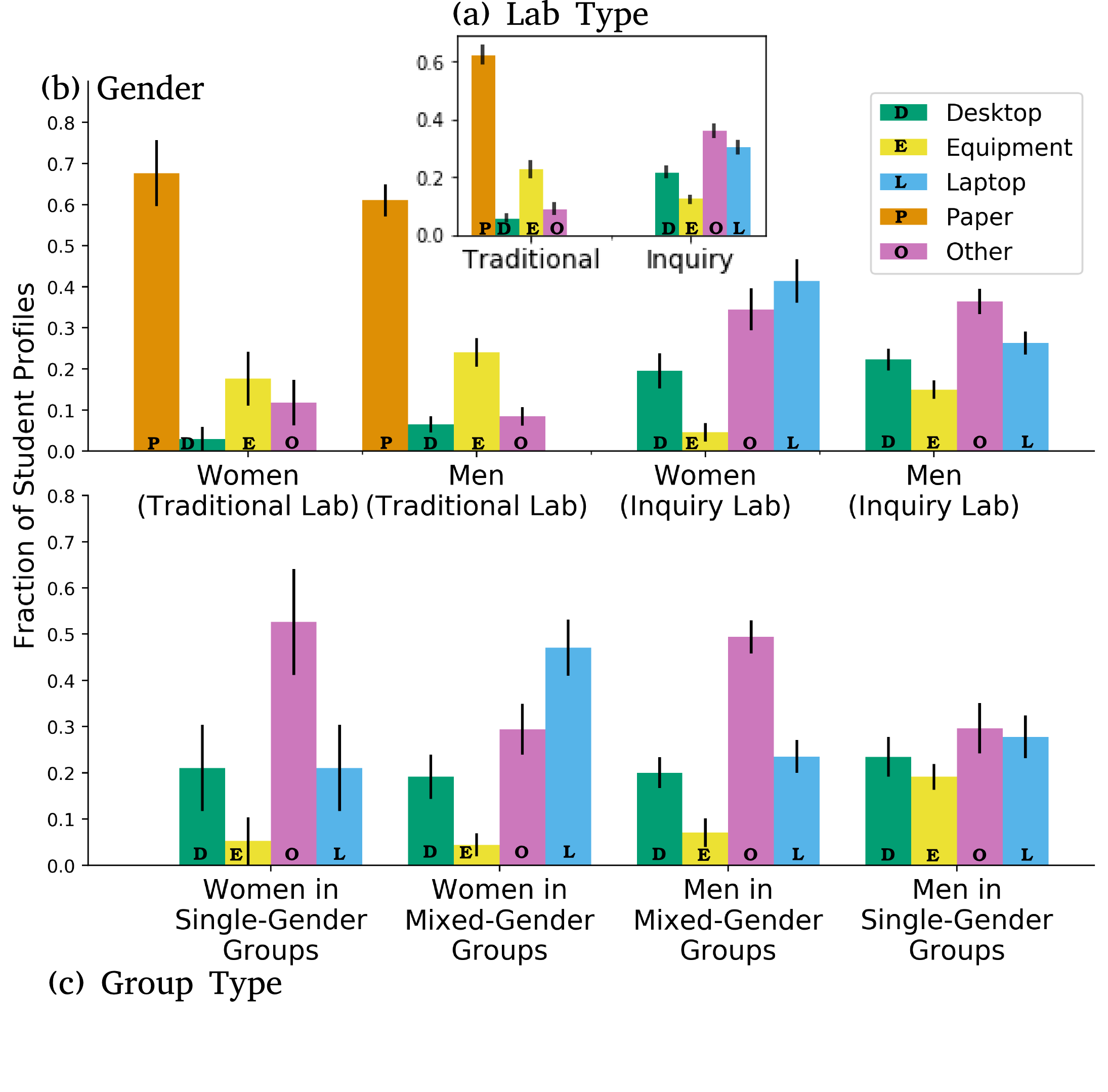}
    \caption{\textbf{Cluster compositions} for each of the five clusters, broken down both by lab type, gender and group composition. In all plots, y-axis represents fraction of student profiles and errors are calculated using the standard error on the fraction of a population shown (see Eq.~\ref{eq:PopErr} for additional details). (a)~Cluster distributions broken down by lab type. (b)~Clusters further broken down by gender. We see that there are disproportionately more women in the Laptop cluster than men, and disproportionately more men than women in the Equipment cluster. (c)~Cluster distributions were further broken down in the inquiry lab by group type (men and women in mixed-gender groups and single-gender groups). Upon inspection, we see that the Laptop difference remained, while a difference emerged in Other. Furthermore, far more men are high-equipment users when in single-gender groups. Due to insufficient statistics, no comparison can be made with women in single-gender groups, and the data are presented for completeness.}
    \label{fig:clusterDistribution}
\end{figure*}

When broken down by lab type (shown in Fig.~\ref{fig:clusterDistribution}{(a)}), \smash{$60\%$} of the student profiles in the traditional labs were in the Paper cluster, indicating that the majority of students in the traditional labs were high paper users. Students in the inquiry labs engaged in a more varied set of activities, demonstrated by the uniform distribution of student profiles across clusters. In the traditional labs, however, student profiles were predominantly found in the Paper cluster, with few profiles in the remaining clusters.

\begin{figure}
    \centering
    \includegraphics[width=0.45\textwidth]{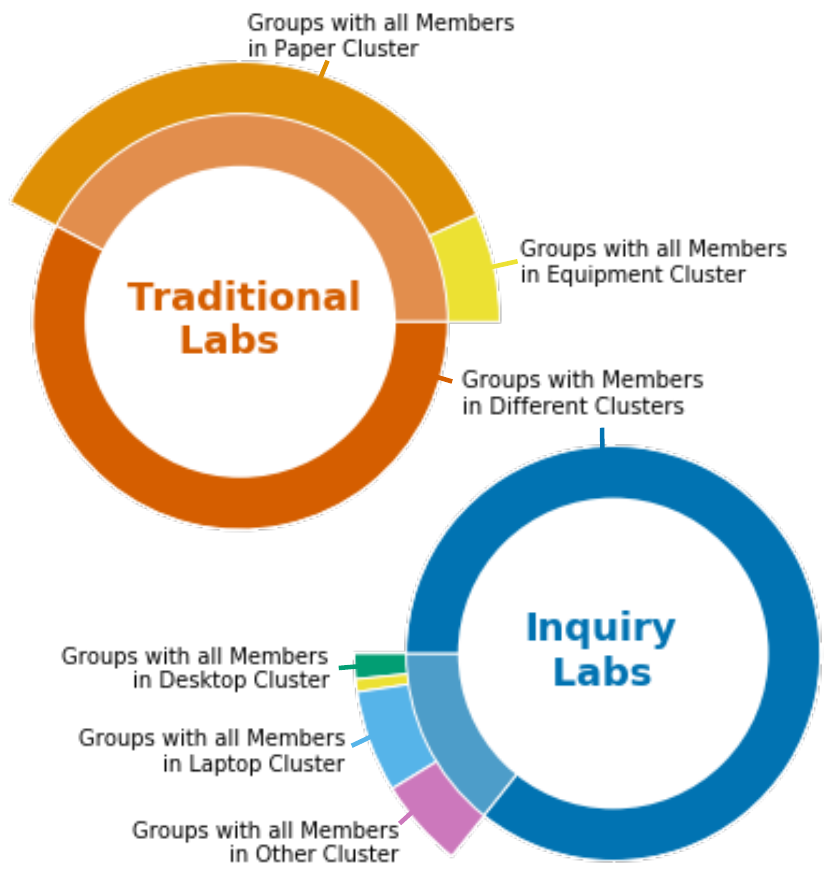}
    \caption{\textbf{Fraction of groups} with members in identical clusters (light ring) and different clusters (dark ring) illustrating role division in the different labs. Almost half of groups in the traditional labs had all members in the same cluster (primarily Paper cluster), whereas the majority of groups in the inquiry labs had members in multiple clusters indicating an increase in task division.}
    \label{fig:groupClusters}
\end{figure}

Our data support the notion that labs with reduced structure provide a wider range of available roles. We tested this explanation by examining the range of roles within individual groups in each class type: Do members within a group predominantly fall into the same or different clusters? In the traditional labs, \smash{$43\%$} of groups had all members in the same cluster (predominantly the paper cluster) whereas only \smash{$14\%$} of groups in the inquiry labs had all members in the same cluster (Fig.~\ref{fig:groupClusters}).  

We note that groups in the traditional and inquiry labs were of varying sizes. Groups in the traditional labs typically had three or four students, whereas groups in the inquiry labs typically had two or three members, with group sizes determined by logistical constraints of the lab spaces (such as the number of available lab benches given the size of each class) and mainly assigned randomly by the instructor. Moreover, mixed-gender groups also had between 1-3 women and 1-3 men. Observers documented the behavior of all students in every group, and kept track of which student was in which group. One could expect that, in groups with more members, there is an increased chance of task division occurring. While groups in the traditional labs typically had more members than those in the inquiry labs, Fig.~\ref{fig:groupClusters} in fact shows proportionally fewer groups in the inquiry labs with members in identical clusters, supporting the conclusion that groups in the inquiry labs were more likely to divide tasks.

We infer that the set of available roles is much greater in the inquiry labs and that students assumed distinct roles from one another. The traditional labs were highly guided, leaving students little room for active decision-making about the experiment. While they worked in groups, each student was responsible for completing their own individual worksheet. As a result, the set of available roles was both confined and manifestly similar for all students. In contrast, the inquiry labs were designed to emphasize the process of experimentation and thus students supported in exercising agency for active decision-making about the experiment. As a result, the set of available roles was larger and students could divide tasks in a variety of ways.  

We next sought to evaluate whether men and women assume different roles. We decomposed the behavior clusters by gender and lab type, as shown in Fig.~\ref{fig:clusterDistribution}{(b)}. Through a chi-squared test of frequencies, we found a statistically significant difference between men and women in the inquiry labs ($\chi^2(3) = 10.77$, $p=0.01$, $V_{Cramer} = 0.15$) but none in the traditional labs ($\chi^2(3) = 3.27$, $p=0.65$, $V_{Cramer}=0.08$). There were disproportionately more women in the Laptop cluster than men and disproportionately more men in the Equipment cluster than women. 

Statistically significant differences also existed between men in mixed- versus single-gender groups, shown in Fig.~\ref{fig:clusterDistribution}{(c)} ($\chi^2(3)=12.10$, $p=0.007$, $V_{Cramer}=0.15$). When men were in single-gender groups, they were more likely to be in the Equipment cluster and less likely to be in the Other cluster than men in mixed-gender groups. Men in mixed-gender groups were more likely than their female group members to be in the Other cluster, and women in mixed-gender groups were more likely than their male group members to be in the Laptop cluster ($\chi^2(3)=10.34$, $p=0.02$, $V_{Cramer}=0.15$). Due to the small number of women in single-gender groups, we did not have statistical power to detect whether there are differences for women who were in mixed versus single-gender groups ($p>0.17$ in all cases). Furthermore, due to insufficient statistics, we were unable to perform a similar analysis for groups of varying sizes.

The difference in men's behavior when in mixed- and single-gender groups may be indicative of the impact of social context on the roles students assume. In groups with only men, there may be different social dynamics compared to groups that include women, changing the set of available roles (and thus observed behaviors). For instance, the increased number of high-equipment users in men-only groups may be the result of ``playfulness''~\cite{Hasse2008} when women are not in the group, or that in mixed-gender groups members were more efficient with equipment use. 

\subsection{Quantifying the relative behaviors of students within groups}
\label{sec:intragroupMainText}

The cluster analysis in the previous section indicates that individual students took on different roles on a course-wide level but suggests that group composition may impact the group dynamics in a non-trivial way. To investigate roles within individual groups, and to ensure that different analysis methods obtain non-conflicting results, we compared each student's profile to those of their group members. We quantified the relative behaviors by constructing a \textit{deviating profile} for each student to describe how they differed from their group's average profile (quantified as the numerical difference of the student profile from the group average, see Appendix~\ref{sec:intragroupAppendix} for additional details). For example, if all students in a group behaved the same, the profiles of every student would match their group's average, and thus they would each have a deviation of zero for each code. The distribution of all students' deviations for each code has a mean of zero, as the deviations in every group must cancel each other out. However, the variances of these distributions (defined here as the \textit{intragroup variance}) are not constrained and indicate the degree of task division. An intragroup variance of zero implies that any student's behaviors are completely indistinct from their group, while a large intragroup variance reveals a greater degree of divide-and-conquer.

\begin{figure}
    \centering
    \includegraphics[width=0.3\textwidth]{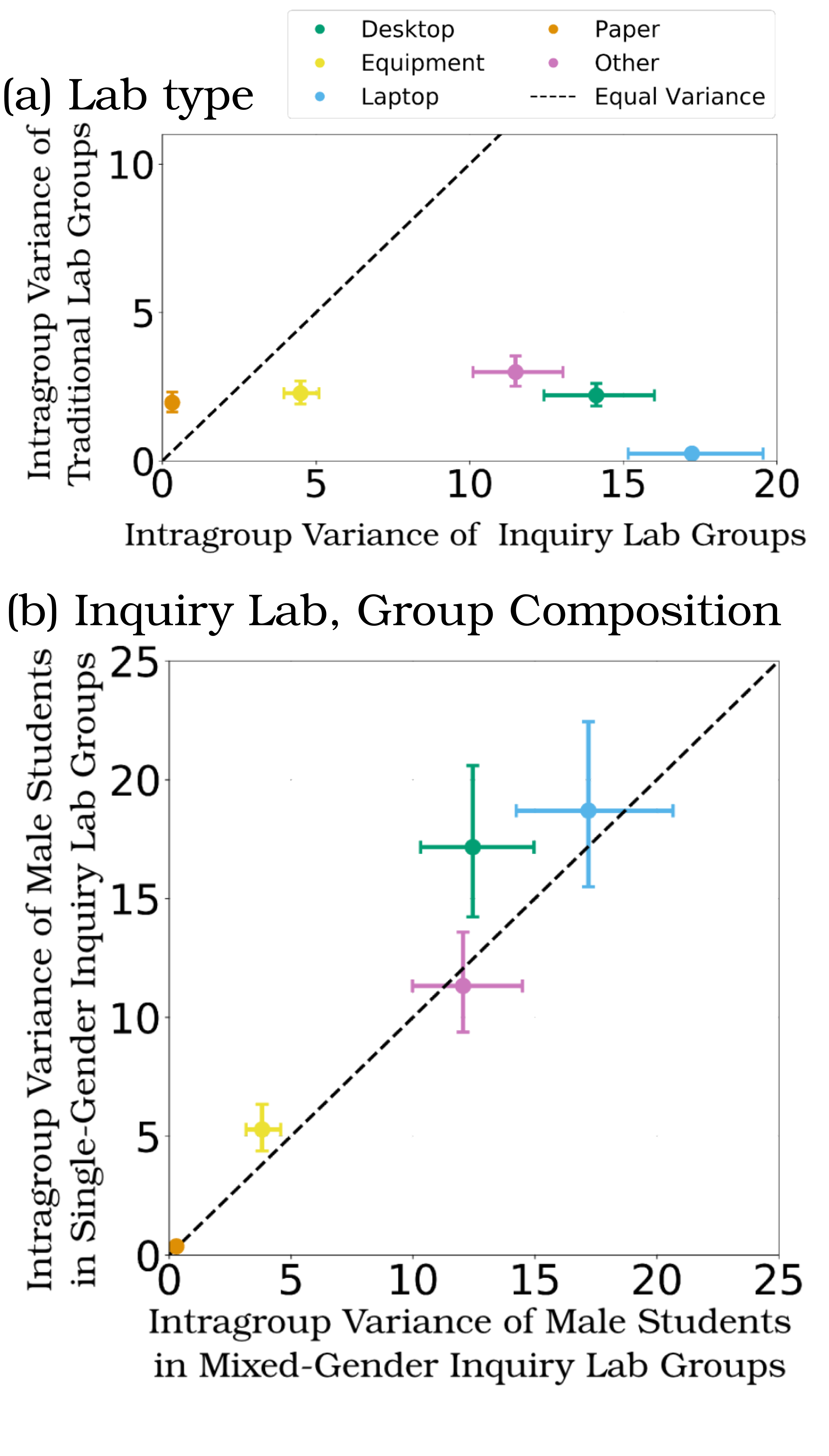}
    \caption{\textbf{Intragroup variances} of the relative behaviors among students, signifying the amount of task division within groups. Each plot shows VAR($\Delta N$) for all student profiles contained within the labeled lab and group types along with their Bayesian confidence intervals. (a)~Comparing across lab types, the intragroup variances are remarkably larger in the inquiry lab groups than in the traditional lab groups for all codes besides paper, indicating a greater range of behaviors and an increase in task division. (b)~Within the inquiry labs, the intragroup variances are comparable among groups of differing composition suggesting that similar degrees of task division were taking place. (Female single-gender groups not included due to insufficient statistics).}
    \label{fig:intragroupVariances}
\end{figure}

In the traditional labs, the intragroup variance was very small for all coded behaviors other than Paper (Fig.~\ref{fig:intragroupVariances}(a)). This result supports the analysis and interpretation from the cluster analysis: groups in the traditional labs did not divide roles and each student behaved similarly to their group members. In the inquiry labs, intragroup variances were much larger for all codes apart from Paper, which indicate a high degree of task division took place.

Within the inquiry labs, we found comparable intragroup variances among all coded behaviors regardless of the group's composition (that is, single- versus mixed-gender groups; Fig.~\ref{fig:intragroupVariances}(b)). This result 
suggests that the group composition does not impact the group dynamics with respect to the amount of task division; that is, single- and mixed-gender groups divide roles to similar degrees. 

However, within mixed-gender groups, these roles are divided along gender lines. The distributions of deviations for men and women in mixed-gender groups differed significantly for the Laptop ($p=0.001$) and Other ($p=0.009$) codes. Women handled a laptop or personal device more than their group members, and men participated in Other activities more than their group members. Furthermore, men in mixed-gender groups appeared to handle equipment more often than the group average, however this result was only marginally significant with the Bonferroni correction ($p=0.012$). See Appendix~\ref{sec:intragroupAppendix} and Fig.~\ref{fig:intragroupDeviations} for supporting data. To better understand these dynamics within mixed-gender groups, we sought a more fine-grained description of the roles students take on and how they are assigned.

\subsection{Understanding specific student tasks and identifying role assignments} 
We captured video of a subset of individual groups for entire lab periods and identified the specific tasks associated with the coarse behaviors discussed in the previous sections. For example, when a student was handling the equipment, were they collecting data or setting up the apparatus? We identified the specific tasks through visual cues and students' speech. We then measured the total amount of time each student spent on each specific task. 

The cluster and intragroup analyses found significant differences between men and women in mixed-gender groups with regards to laptop usage and Other activities. The individual group video analysis found that women spent about twice as much time as men analyzing data on laptops ($14\pm7\%$ of the lab period for women and $6\pm3\%$ for men). However, we did not find a clear difference in the specific tasks associated with the Other behavior between men and women in this subset of groups. The biggest difference among Other tasks came from within-group behaviors such as talking, observing, or interacting with group members ($30\pm4\%$ of the lab period for men and $26\pm5\%$ for women). 

We also used the single-group video analysis to identify that in almost all cases, students did not discuss the roles they would assume. Notably, there were no instances of explicit role allocation from peers in the group or from lab instructors. We conjecture students either self-assigned roles within groups, `fell into' roles, or directed each other through \textit{positioning} (subtle verbal and non-verbal social cues~\cite{Davies1990,Berge2013}). Exploring mechanisms for role allocations is the focus of future study, to better understand how roles become gendered. Tentatively, we conclude that the significant difference in roles is not the result of overt, explicit allocation. Rather, we infer that subtle interactions at the individual level accumulate to create class-level patterns. 

\section{Discussion and Conclusions}
In this study, we identified how student behaviors in a lab vary by lab type, gender, and group composition. From coarse-grained observations of what students were handling in the lab, we found that students in traditional labs generally behave similarly, spending most time writing on the lab worksheets. Behaviors in the inquiry labs were much more varied, with behaviors focused on using equipment and computers. Furthermore, women in the inquiry labs tended to be high laptop users (primarily analyzing data), while men were high equipment users (collecting data or manipulating the equipment). This pattern varied by group composition, however, where men in mixed-gender groups were much more often engaged in Other behaviors (primarily talking to their peers), while men in single-gender groups were the high equipment users. Within-group analyses indicated that these differences were a result of group members taking on distinct roles, rather than whole groups tending towards similar behaviors. The role division was not a result of explicit allocation between group members. 

Research indicates that providing students with more authentic lab experiences, often by removing structure to grant students more agency, improves student attitudes towards science and engagement in high-level scientific practices~\cite{WilcoxGuided, Holmes2015, Etkina2010, Brownell2012, Adams2009}. The results here suggest that by removing structure in labs, these curricula facilitate student-driven group work and open up a new set of group roles, but may unintentionally create inequitable learning environments or provide the opportunity for underlying inequities to manifest. Increased student-agency, on its own, is insufficient for the creation of a supportive and equitable learning environment, where each student has the opportunity to freely pursue their own path in physics. Equitable participation must be actively built into curricula, to eliminate implicit inequities that can go on behind the scenes.

We have found that inquiry-based labs, designed to support student decision-making, increased the variation in student behaviors when compared to the more traditional lab structure. Working collectively in groups, with a pedagogical structure that facilitated group work (such as having one electronic notebook per group as opposed to identical, individual worksheets) opened up new group roles and increased the range of behaviors students took on. Removing structure in lab activities so that students may take on a variety of roles supports a variety of students experiences during an activity. Through these experiences, we may communicate to students that there are multiple ways to contribute to science and to be a physicist.

However, the freedom for students to fall into roles without any guidance or pedagogical structure has the potential to introduce problematic inequities. While one could argue that allowing students to assume the roles they are more comfortable with may increase persistence in the course (regardless of whether or not they are gendered), we note that in the absence of structuring equitable participation and group work students may inadvertently fall back on cultural norms and expectations when taking on roles within their group and may rely on implicit biases when making these decisions. Each student's experience is unique in a classroom, but systematic differences in these experiences may have unintended, detrimental consequences. In this study, systematic gendered inequities (with men and women systematically taking on different group roles) and group behavior that depends on group composition (men behaving differently when in groups with other men versus when there is at least one woman) were statistically apparent only in a curriculum that provided ample agency. If such differences are supported in institutional settings, they can contribute to increased gender segregation through students' educational experience, and ultimately contribute to the large gender imbalance seen in the field as a whole.

The focus of this study was intentionally directed at students primarily intending to major in physics. While this narrow population limits generalization to non-physics majors, it provides vital information with regards to group work, which has potential implications on students' identities as physicists and decisions to persist in physics~\cite{Carlone2007}. This work also has implications for instruction. Our data, however, do not speak to the efficacy of different approaches at mitigating the issues observed. We can draw from previous literature to propose strategies that should be studied. For example, it has been shown that increased pedagogical structure combined with active learning can reduce the achievement gap in class work~\cite{Haak1213}. Therefore, actively building into lab curricula group roles~(similar to those of cooperative grouping~\cite{Heller1992}) such as ``group PI'', ``reviewer'' or ``science communicator'' that have students actively think about how roles are assigned and make deliberate choices regarding role division could alleviate the unintended consequences of subconsciously acting on implicit biases, and is the focus of further research. 

Previous work has identified many structural manipulations that support equitable participation in other learning environments~\cite[e.g.][]{Tanner2013, Heller1992}. Our results highlight that there may be unique challenges to equity in inquiry lab environments, where students divide roles associated with distinct experimentation tasks (such as analyzing data or handling equipment). The existence of role division is not inherently problematic. However, the different roles physics students take on can greatly influence their unique experience, identity formation, and sense of belonging, which, in turn, ultimately impact persistence and representation in the field~\cite{Carlone2007, Rainey2018, Fisher2019}. With many calls to reform lab instruction to provide students with more authentic experiences and less structure, researchers have a responsibility to evaluate the potential side effects of such interventions. Given the many issues in representation and persistence in STEM, students' experiences should not be sacrificed for the increased learning benefits of these kinds of labs. Instructors have the responsibility of ensuring that the desired aspects of research and academia are being reinforced in these learning environments, and that we are not inadvertently reinforcing gendered roles by failing to actively intervene.

\appendix

\section{Statistical Analysis of Intragroup Variances}
\label{sec:intragroupAppendix}

Here we present our intragroup analysis procedure to investigate whether roles emerged within individual groups. Each lab period involved groups of students working together as a team to progress through an experiment. We compared each student profile in a group to their group's average profile and quantified how a student deviated from their group's average for each code. Rescaling each profile in a group with respect to that group's average reveals the variations between the group-members' behaviors. We then compared whether there were any significant differences between the relative behaviors of men and women.

We quantified the relative behaviors of students by constructing each student's \textit{deviating profile}. If the coded behaviors were distributed equally within a group, then the observations of each student would match the group's average for each code. Denoting the observed count of a coarse behavior code for student $S$ in group $G$ with $N_\textrm{code}^S$, the expected count of that code for a student in that group is:
\begin{equation}
 \langle N_\textrm{code} \rangle_G = \frac{1}{M_G}\sum_{S \in G}^{M_G} N_\textrm{code}^S,
\end{equation}
where the sum runs over each student in one group with $M_G$ total group members. From this expectation value, we calculate how student $S$ deviates from their group's average
\begin{equation}
\label{eq:deltaN}
\Delta N_\textrm{code}^S = N_\textrm{code}^S-\langle N_\textrm{code} \rangle_G.
\end{equation}
These deviations reveal interesting behavior trends within groups. For instance, a student engaging in a particular task more than their group members would be revealed with a large and positive $\Delta N$.

The distribution of deviations $\Delta N$ for each code provides information about task division within groups. When the distribution of deviations contains all group members, the mean is constrained to zero since each student's deviation cancel each other out by definition. However, the variance of these distributions for each code (the \textit{intragroup variance} defined as VAR($\Delta N$)) are not constrained and provide a measure of the amount of task division within a group. Zero variance among deviations would imply the students' behaviors are completely indistinct from another while a large variance would reveal a greater degree of divide-and-conquer. 

\begin{figure}
    \centering
    \includegraphics[width=0.5\textwidth]{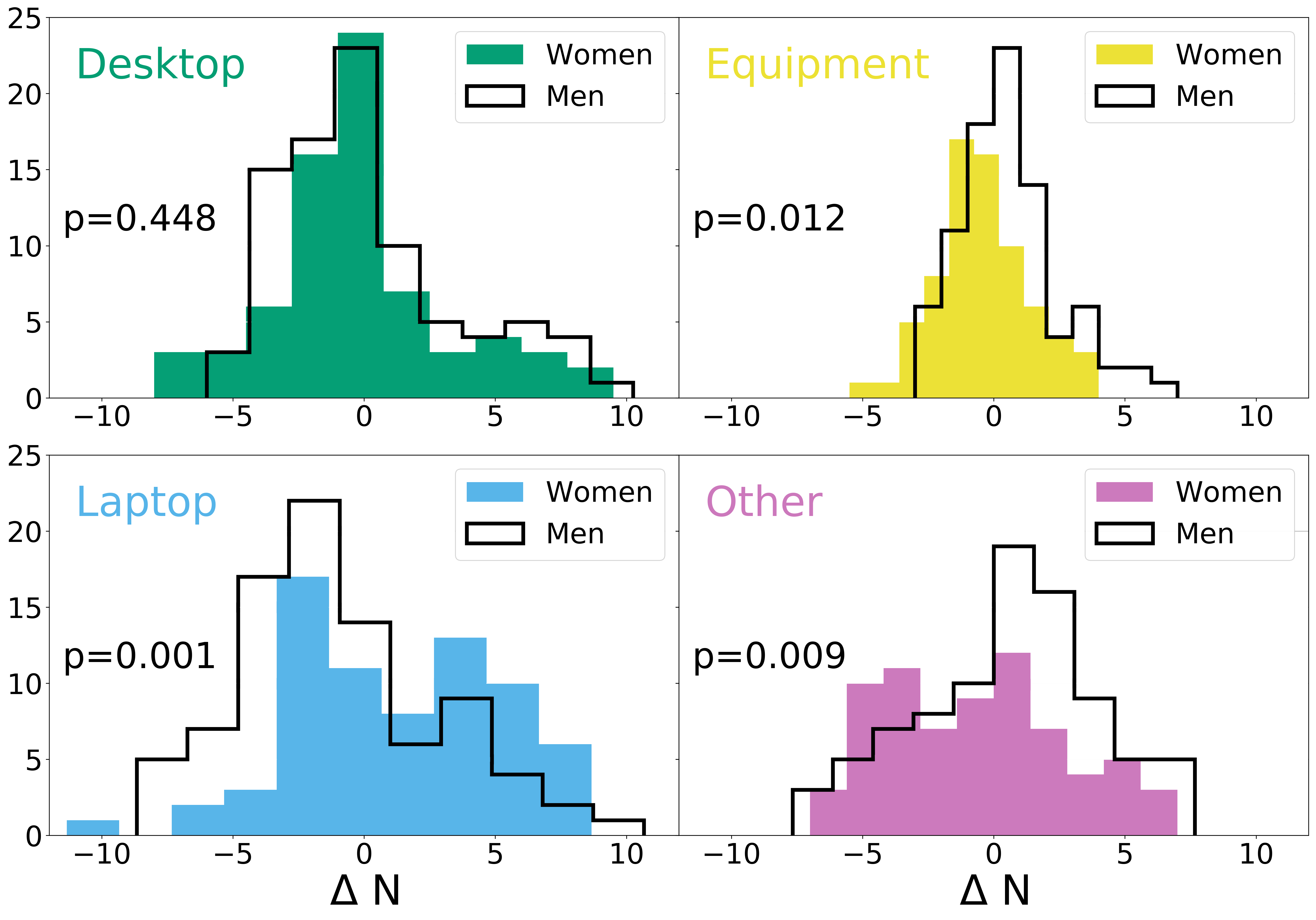}
    \caption{\textbf{Histograms of intragroup deviations} for men and women within the inquiry lab's mixed-gender groups for the Desktop, Equipment, Laptop, and Other behavior codes, with y-axis representing the number of student profiles. Each student deviates from their group's average by $\Delta N$ (defined in Eq.~\ref{eq:deltaN} in Materials and Methods~\ref{sec:intragroupAppendix}). A positive $\Delta N$ denotes a student engaging in a behavior more often relative to their group members. We quote $p$-values calculated from the Mann-Whitney~U test statistic on all plots and find significant differences between men and women for the Laptop and Other codes. We also find a borderline result of men handling the equipment more than their group members.}
    \label{fig:intragroupDeviations}
\end{figure}

In Fig.~\ref{fig:intragroupVariances}(a), we plot the intragroup variances for the two lab types. The relative behaviors within groups from the inquiry labs were highly varied when compared to the traditional labs, which exhibited remarkably less variance for all codes except Paper. This result was confirmed with Levene's test to assess the equality of variances, where none of the $p$-values from the test statistic for each code exceeded $10^{-5}$. This disparity among intragroup variances was expected as the traditional labs were highly guided and students were required to fill out their own worksheet, while the inquiry labs were less guided and students were given more agency for active decision-making about the experiment. The large intragroup variances in the inquiry lab groups signify a higher degree of task division taking place.

To investigate task division in the inquiry lab groups, we compared the intragroup variances for different group compositions (Fig.~\ref{fig:intragroupVariances}(b)). We found comparable intragroup variances regardless of the group's composition for all behavior codes (every code's $p$-value from Levene's test exceeding the $p=0.01$ cutoff, with $p$-values ranging from $p=0.03-0.9$). The comparable intragroup variances signify that there was no significant difference in the degree of task division in the inquiry lab groups with respect to group composition. 

To examine the relative behaviors of men and women, we shifted our focus to within mixed-gender groups. In Fig.~\ref{fig:intragroupDeviations}, we plot the histograms of deviations for men and women in mixed-gender groups for the Desktop, Equipment, Laptop, and Other codes. We performed a Mann-Whitney~U test as a non-parametric test to determine whether there were any significant differences among the distributions from men and women. We find significant differences in the deviations from men and women for the Laptop ($p=0.001$) and Other ($p=0.009$) codes. Women handled a laptop or personal device more than their group members, and men participated in Other activities more than their group members. We also find that men in mixed-gender groups appear to handle equipment more often than the group average ($p=0.012$), however this result was only marginally significant with the Bonferroni correction.

\begin{acknowledgements}We thank the teaching assistants and lab instructors for the course used in this study for their invaluable support and cooperation. We also thank Chris Gosling for valuable conversations and insight, and James Sethna for helpful feedback. This material is based upon work supported by the National Science Foundation under Grant No. 1836617, the President's Council for Cornell Women's Affinito-Stewart Grant, and the Cornell University College of Arts and Sciences Active Learning Initiative.
\end{acknowledgements}

\bibliography{GiL}

\end{document}